\documentclass[10pt]{article}
\usepackage{geometry}
\usepackage{amsmath,amssymb}
\usepackage{xcolor}
\definecolor{graycolor}{gray}{0.9} 
\usepackage{microtype}
\usepackage{setspace} 
\usepackage[utf8]{inputenc}
\usepackage[english]{babel}
\usepackage{times}
\usepackage{array}
\usepackage{soul}
\usepackage{cite}
\usepackage[numbers,sort&compress]{natbib}
\usepackage{natbib}

\usepackage{titlesec} 
\titleformat {\section} [block] {\raggedright \fontsize{10}{10}\selectfont\bfseries} {\thesection. \space} {0pt} {}
\titlespacing {\section} {0pt} {12pt} {6pt}
\titleformat {\subsection} [block] {\raggedright \fontsize{10}{10}\selectfont\itshape} {\thesubsection .\space} {0pt} {}
\titlespacing {\subsection} {0pt} {12pt} {6pt}
\titleformat {\subsubsection} [block] {\raggedright \fontsize{10}{10}\selectfont} {\thesubsubsection .\space} {0pt} {}
\titlespacing {\subsubsection} {0pt} {12pt} {6pt}
\titleformat {\paragraph} [block] {\raggedright \fontsize{10}{10}\selectfont} {} {0pt} {}
\titlespacing {\paragraph} {0pt} {12pt} {6pt}

\usepackage{array} \newcommand{\PreserveBackslash}[1]{\let\temp=\\#1\let\\=\temp}
\newcolumntype{C}[1]{>{\PreserveBackslash\centering}m{#1}}
\newcolumntype{R}[1]{>{\PreserveBackslash\raggedleft}m{#1}}
\newcolumntype{L}[1]{>{\PreserveBackslash\raggedright}m{#1}}
\usepackage{tabularx}
\usepackage{colortbl}
\usepackage{graphicx}
\usepackage{float}
\usepackage[export]{adjustbox}
\usepackage{caption}
\usepackage{fancyhdr} 
\usepackage{lastpage}
\usepackage{layout}
\usepackage{setspace} 
\usepackage{enumitem}
\usepackage{booktabs}
\usepackage{arydshln}
\usepackage{multirow}
\usepackage{color}
\usepackage{hyperref} 
\hypersetup{
	colorlinks=true,
	linkcolor=blue,
	filecolor=blue,
	urlcolor=black,
	citecolor=cyan,
}
\setcitestyle{open={[},close={]},citesep={,\!},numbers}

\fancypagestyle{firstpage}{
    \setlength{\headsep}{2.2cm}
    
    \setlength{\footskip}{1.5cm}
    \fancyhf{}
    \lhead{\begin{table}[H]
        \centering
        \begin{tabular}{L{2.5cm}C{10cm}C{3.1cm}R{2cm}}
            \includegraphics[scale=0.035]{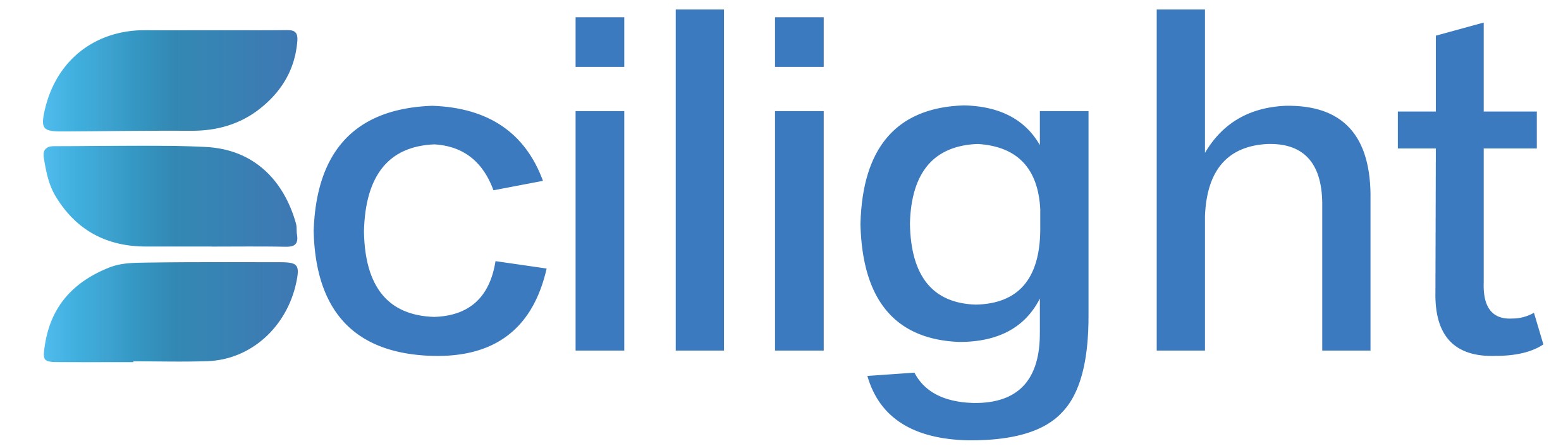} \vspace{-6pt}& \cellcolor{graycolor}\begin{tabular}[c]{@{}c@{}}\textit{International Journal of Gravitation and Theoretical Physics}\\ \href{https://www.sciltp.com/journals/ijgtp}{https://www.sciltp.com/journals/ijgtp}\end{tabular} & \includegraphics[scale=0.014]{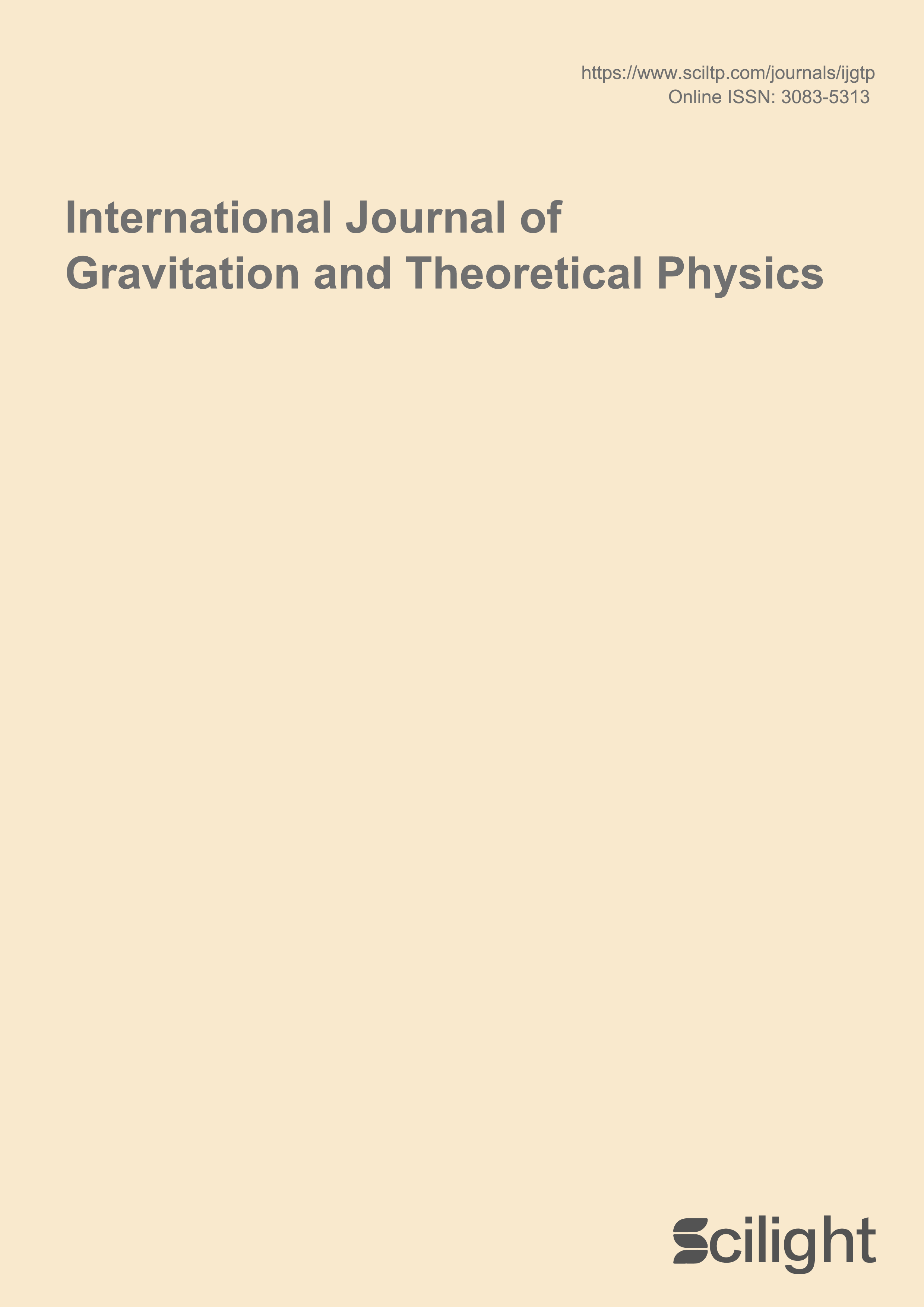} \vspace{-3pt}\\
        \end{tabular}
        \vspace{-22pt}
    \end{table}}
   
    \fancyfoot[C]{
        \vspace{-1.55cm}
        \begin{table}[H]
            \begin{minipage}[c]{0.15\columnwidth}
                \includegraphics[scale=0.5]{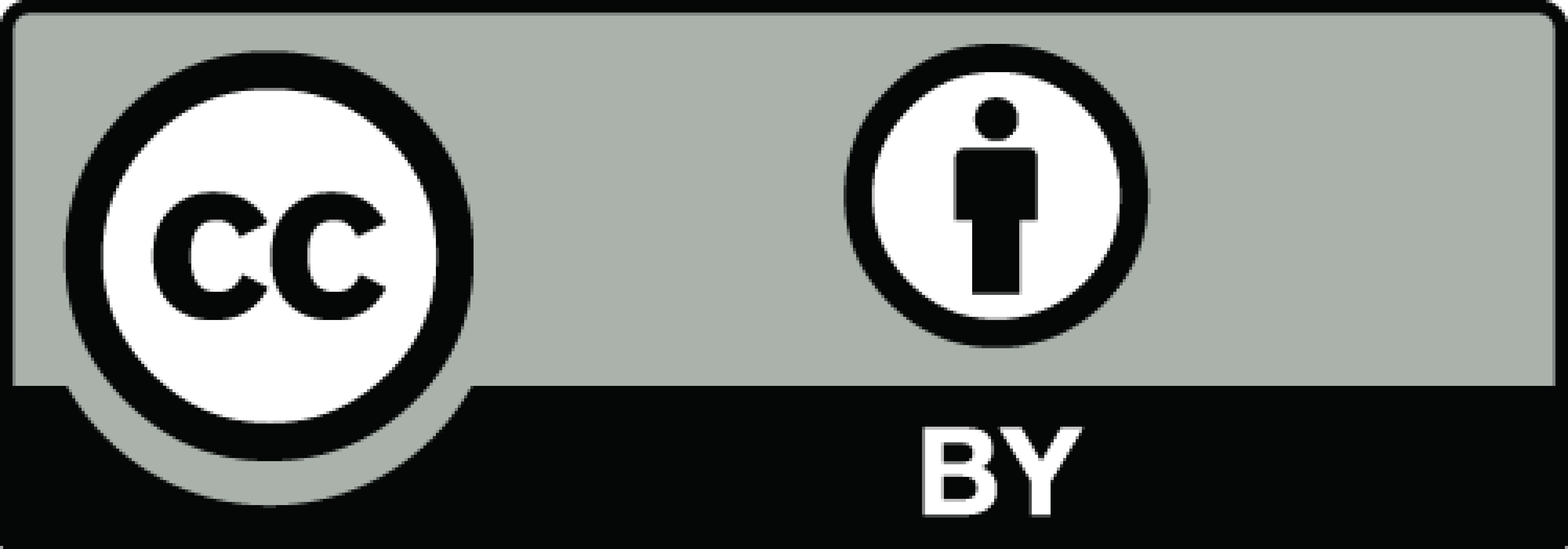} \vspace{1.1pt}
            \end{minipage}
            \hfill
            \begin{minipage}[c]{0.85\columnwidth}
                \scriptsize \textbf{Copyright:} © 2026 by the authors. This is an open access article under the terms and conditions of the Creative Commons Attribution (\mbox{CC BY}) license (\href{https://creativecommons.org/licenses/by/4.0/}{https://creativecommons.org/licenses/by/4.0/}). \\ \textbf{Publisher’s Note:} Scilight stays neutral with regard to jurisdictional claims in published maps and institutional affiliations.
            \end{minipage}
    \end{table}}
    \vspace{-0.55cm}
}

\begin{document}
\newgeometry{left=2.5cm, right=2.5cm, top=1.8cm, bottom=4cm}
	\thispagestyle{firstpage}
	{\noindent \textit{Article}}
	\vspace{4pt} \\
	{\fontsize{18pt}{10pt}\textbf{Orbital, Shadow and Thin-Disk Signatures of a Regular Black Hole with Gravitational Self-Energy}}
	\vspace{16pt} \\
	{\large Erdinç Ulaş Saka}
	\vspace{6pt}
	 \begin{spacing}{0.9}
		{\noindent \small
			Department of Physics, Faculty of Science, Istanbul University, Vezneciler, Istanbul 34134, Türkiye; ulassaka@istanbul.edu.tr \vspace{6pt}\\
		\footnotesize	\textbf{How To Cite}: Saka, E.U. Orbital, Shadow and Thin-Disk Signatures of a Regular Black Hole with Gravitational Self-Energy. \emph{International Journal of Gravitation and Theoretical Physics} \textbf{2026}, \emph{2}(2), 5. \href{https://doi.org/10.53941/ijgtp.2026.200005}{https://doi.org/10.53941/ijgtp.2026.200005}}\\
	\end{spacing}

\begin{table}[H]
\noindent\rule[0.15\baselineskip]{\textwidth}{0.5pt} 
\begin{tabular}{lp{12cm}}  
 \small 
  \begin{tabular}[t]{@{}l@{}} 
  \footnotesize  Received: 23 May 2026 \\
  \footnotesize  Revised: 25 June 2026 \\
   \footnotesize Accepted: 26 June 2026 \\
  \footnotesize  Published: 29 June 2026
  \end{tabular} &
  \textbf{Abstract:} We investigate geodesic motion, shadow observables and thin-disk accretion for the regular black hole generated by a non-local gravitational self-energy contribution. The geometry is controlled by a zero-point length and can be followed smoothly from the Schwarzschild limit to a cold extremal remnant. We compute the photon ring, critical impact parameter, apparent shadow radius, null Lyapunov exponent, innermost stable circular orbit, orbital frequencies and Novikov-Thorne flux profile. As the self-energy scale grows, both the photon ring and the ISCO move outward, while the photon-ring frequency and instability exponent decrease. The horizon-normalized shadow area increases by about a factor of 2.7 for the near-extremal benchmark, although the same shadow radius decreases when normalized by the ADM mass. The ISCO binding efficiency grows modestly, whereas the zero-torque thin-disk flux peak moves outward and falls to about 61\% of the Schwarzschild peak as the solution approaches the remnant regime. These trends identify a coherent set of optical, orbital and accretion signatures of the gravitational-self-energy regularization.\\
\\
  & 
\end{tabular}
\noindent\rule[0.15\baselineskip]{\textwidth}{0.5pt} 
\end{table}

	\section{Introduction}

Regular black holes provide a useful phenomenological way to parametrize short-distance departures from the Schwarzschild solution without introducing a curvature singularity. They are especially intriguing because they connect several open questions at once: the possible resolution of classical singularities, the existence of compact remnants, quantum-gravity-inspired modifications of the near-core geometry, and potentially observable changes in lensing, shadows, accretion and wave dynamics \cite{Bardeen1968,Hayward2006,Konoplya:2025ect,Nicolini:2005vd,Spina:2025wxb,Ayon-Beato:1998hmi}. The solution considered here was introduced by Jusu\'{f}i and Singleton as a neutral black hole sourced by a regularized gravitational self-energy in a T-duality-inspired non-local framework \cite{JusufiSingleton2026}. Here $l_0$ represents a zero-point length, namely an effective minimal resolvable distance that smears the central source and sets the scale at which the classical Schwarzschild geometry is regularized. The geometry reduces to Schwarzschild when $l_0\to0$, and ends at a zero-temperature extremal remnant when the horizon radius becomes comparable to $l_0$.

The same deformation that changes the near-core structure also changes particle motion outside the event horizon. Circular null geodesics determine the photon ring, the high-frequency capture cross section and the apparent shadow radius seen by an observer at infinity \cite{Synge1966,Bardeen1973,Luminet1979,Falcke2000,EHT2019M87I}, while their Lyapunov exponents govern the instability time scale of the light ring and enter the eikonal connection with quasinormal ringing \cite{Chandrasekhar1983,Cardoso2009Geodesic}. Timelike circular geodesics determine the ISCO, the binding energy available to an accretion disk, and the characteristic orbital frequencies measured by distant observers \cite{BardeenPressTeukolsky1972}. For thin, radiatively efficient disks, the same circular-orbit data feed directly into the Novikov-Thorne/Page-Thorne flux profile \cite{NovikovThorne1973,PageThorne1974}.

\textls[10]{This paper focuses on conservative particle dynamics, shadow observables and disk diagnostics. We keep the outer horizon fixed, write all radii in units of $r_+$, and follow the black-hole branch through the benchmark values $\lambda=0,0.2,0.4,0.6,0.65$, with the final point close to $\lambda_{\rm ext}\simeq0.6526$. The main qualitative result is simple: increasing $l_0/r_+$ moves characteristic circular orbits outward, enlarges the horizon-normalized shadow, lowers the photon-ring frequency, reduces the ISCO frequency near the remnant branch, and slightly increases the ISCO}


\noindent binding efficiency. The same shadow radius decreases when normalized by $M_{\rm ADM}$, showing that the observational interpretation depends on which mass or length scale is held fixed.

The paper is organized as follows. Section~\ref{sec2} introduces the horizon-normalized form of the metric and the black-hole branch. Section~\ref{sec3} summarizes the null and timelike geodesic relations used throughout the analysis. Sections~\ref{sec4} and \ref{sec5} discuss photon rings, null instability and shadow-radius diagnostics. Section~\ref{sec6} studies timelike circular orbits and the ISCO, while Section \ref{sec7} applies the same orbital data to a zero-torque Novikov-Thorne thin disk. Section~\ref{sec8} summarizes the main optical, orbital and accretion signatures.

\section{Geometry and Horizon-Normalized Branch}\label{sec2}

The spacetime is static and spherically symmetric,
\begin{equation}
\label{eq:metric}
 ds^2=-f(r)\,dt^2+\frac{dr^2}{f(r)}+r^2d\Omega_2^2 .
\end{equation}

The lapse function of Ref.~\cite{JusufiSingleton2026} is
\begin{equation}
\label{eq:lapse_dimful_particle}
 f(r)=1-\frac{2m r^2}{(r^2+l_0^2)^{3/2}}
 +\frac{m^2r^2}{(r^2+l_0^2)^2}\,F(r),
\end{equation}
where
\begin{equation}
\label{eq:F_dimful_particle}
 F(r)=\frac{5}{8}+\frac{3l_0^2}{8r^2}
 -\frac{3(r^2+l_0^2)^2}{8l_0r^3}\arctan\!\left(\frac{r}{l_0}\right).
\end{equation}

At large radius,
\begin{equation}
\label{eq:adm_mass_particle}
 f(r)=1-\frac{2M_{\rm ADM}}{r}+O(r^{-2}),
 \qquad
 M_{\rm ADM}=m\left(1+\frac{3\pi m}{32l_0}\right).
\end{equation}

We use the horizon-normalized variables
\begin{equation}
 x=\frac{r}{r_+},\qquad
 \lambda=\frac{l_0}{r_+},\qquad
 \bar m=\frac{m}{r_+}.
\end{equation}

Here and below, subscripts $x$ denote derivatives with respect to the dimensionless radius $x$ at fixed $\lambda$. Unless explicitly stated otherwise, all radii are quoted in units of $r_+$ and all frequencies are quoted in units of $r_+^{-1}$.
Then
\begin{equation}
\label{eq:lapse_dimensionless_particle}
 f(x)=1-\frac{2\bar m x^2}{(x^2+\lambda^2)^{3/2}}
 +\frac{\bar m^2x^2}{(x^2+\lambda^2)^2}\,F_\lambda(x),
\end{equation}
with
\begin{equation}
\label{eq:F_dimensionless_particle}
 F_\lambda(x)=\frac{5}{8}+\frac{3\lambda^2}{8x^2}
 -\frac{3(x^2+\lambda^2)^2}{8\lambda x^3}\arctan\!\left(\frac{x}{\lambda}\right).
\end{equation}

The condition $f(1)=0$ fixes $\bar m$ on the black-hole branch. With
\begin{equation}
 A_\lambda=(1+\lambda^2)^{-3/2},\qquad
 B_\lambda=\frac{F_\lambda(1)}{(1+\lambda^2)^2},
\end{equation}
the branch continuously connected to Schwarzschild is
\begin{equation}
\label{eq:mass_branch_particle}
 \bar m(\lambda)=\frac{A_\lambda-\sqrt{A_\lambda^2-B_\lambda}}{B_\lambda},
 \qquad \bar m(0)=\frac12 .
\end{equation}

The endpoint is fixed by the degeneracy conditions
\begin{equation}
 f(1;\lambda_{\rm ext})=0,
 \qquad
 \left.\frac{df}{dx}\right|_{x=1,\lambda=\lambda_{\rm ext}}=0,
\end{equation}
which give
\begin{equation}
 \lambda_{\rm ext}=0.6526091837\ldots .
\end{equation}

The background parameters used below are listed in Table~\ref{tab:metric_parameters_particle}. The inner horizon approaches the outer horizon as $\lambda$ approaches the extremal value, while the Hawking temperature $r_+T_H=f_x(1)/(4\pi)$ tends to zero. At the added near-extremal benchmark $\lambda=0.65$, one has $x_-=0.992025$ and $r_+T_H=3.97\times10^{-4}$. This fixed-horizon normalization is convenient for comparing optical and orbital structures; when a comparison at fixed asymptotic mass is needed, we explicitly use the ADM-normalized quantities.

\vspace{-12pt}

\begin{table}[H]
\centering
\caption{Horizon-normalized background parameters for the black-hole branch. Here $x_-=r_-/r_+$ is the inner horizon when present, and $r_+T_H=f_x(1)/(4\pi)$.}
\label{tab:metric_parameters_particle}
\begin{tabular}{m{1in}<{\centering}m{1.17in}<{\centering}m{1.1in}<{\centering}m{1.1in}<{\centering}m{1.1in}<{\centering}}
\toprule
$\boldsymbol{\lambda}$ & $\boldsymbol{\bar m}$ & $\boldsymbol{M_{\rm ADM}/r_+}$ & $\boldsymbol{x_-}$ & $\boldsymbol{r_+T_H}$ \\
\midrule
        0.0 & 0.500000 & 0.500000 & -- & 0.079577 \\
        0.2 & 0.379121 & 0.590784 & 0.129555 & 0.062845 \\
        0.4 & 0.521561 & 0.721857 & 0.401549 & 0.038216 \\
        0.6 & 0.701936 & 0.943797 & 0.847151 & 0.008054 \\
        0.65 & 0.757500 & 1.017500 & 0.992025 & 0.000397 \\
        \bottomrule
\end{tabular}
\end{table}

\section{Geodesic Equations}\label{sec3}

On the equatorial plane, the conserved energy and angular momentum per unit rest mass are
\begin{equation}
 E=f\dot t,
 \qquad
 L=r^2\dot\phi,
\end{equation}
where a dot denotes differentiation with respect to proper time for massive particles and an affine parameter for null particles. For null rays the overall normalization of $E$ and $L$ is arbitrary, and the invariant impact parameter is $b=L/E$. The radial equation can be written as
\begin{equation}
\label{eq:radial_equation_particle}
 \dot r^2=E^2-V_\epsilon(r),
 \qquad
 V_\epsilon(r)=f(r)\left(\epsilon+\frac{L^2}{r^2}\right),
\end{equation}
with $\epsilon=1$ for timelike geodesics and $\epsilon=0$ for null geodesics.

For timelike circular orbits, $V_1=E^2$ and $V_1'=0$. In the dimensionless coordinate $x=r/r_+$ this gives
\begin{equation}
\label{eq:circular_EL_particle}
 E^2=\frac{2f^2}{2f-xf_x},
 \qquad
 \left(\frac{L}{r_+}\right)^2=\frac{x^3f_x}{2f-xf_x},
\end{equation}
where $f$, $f_x$ and later $f_{xx}$ are evaluated at the circular-orbit radius. The orbital frequency seen at infinity is
\begin{equation}
\label{eq:orbital_frequency_particle}
 (r_+\Omega)^2=\frac{f_x}{2x} .
\end{equation}

The coordinate-time radial epicyclic frequency is
\begin{equation}
\label{eq:epicyclic_frequency_particle}
 (r_+\kappa)^2=\frac{1}{2}\left(ff_{xx}-2f_x^2+\frac{3ff_x}{x}\right).
\end{equation}

The ISCO is the outermost root of $\kappa^2=0$ outside the photon orbit.

Null circular orbits obey
\begin{equation}
\label{eq:photon_condition_particle}
 x f_x-2f=0.
\end{equation}

At the photon ring,
\begin{equation}
\label{eq:photon_observables_particle}
 \frac{b_{\rm ph}}{r_+}=\frac{x_{\rm ph}}{\sqrt{f(x_{\rm ph})}},
 \qquad
 r_+\Omega_{\rm ph}=\frac{\sqrt{f(x_{\rm ph})}}{x_{\rm ph}},
\end{equation}
and the null-orbit instability scale is
\begin{equation}
\label{eq:lyapunov_particle}
 (r_+\Lambda_{\rm ph})^2=
 \frac{f(x_{\rm ph})\left[2f(x_{\rm ph})-x_{\rm ph}^2 f_{xx}(x_{\rm ph})\right]}{2x_{\rm ph}^2} .
\end{equation}

For an observer at spatial infinity in this asymptotically flat spacetime, the apparent shadow radius equals the critical impact parameter,
\begin{equation}
\label{eq:shadow_radius_particle}
 R_{\rm sh}=b_{\rm ph}.
\end{equation}

This identification follows because null rays reaching an observer at infinity are labeled on the observer's image plane by their impact parameter $b=L/E$. The unstable photon orbit gives the separatrix between captured rays and rays that escape or scatter back to infinity. Hence the boundary of the dark region is the critical ray with $b=b_{\rm ph}$, and its apparent radius is $R_{\rm sh}=b_{\rm ph}$ \cite{Synge1966,Bardeen1973,Luminet1979,Falcke2000,EHT2019M87I}.
We quote both $R_{\rm sh}/r_+$ and $R_{\rm sh}/M_{\rm ADM}$ below. The first isolates the change in the near-horizon optical geometry at fixed event-horizon radius, while the second is closer to the normalization used when an asymptotic mass is inferred independently.

\section{Photon Ring and Null Instability}\label{sec4}

\begin{figure}[H]
\centering
\includegraphics[width=\textwidth]{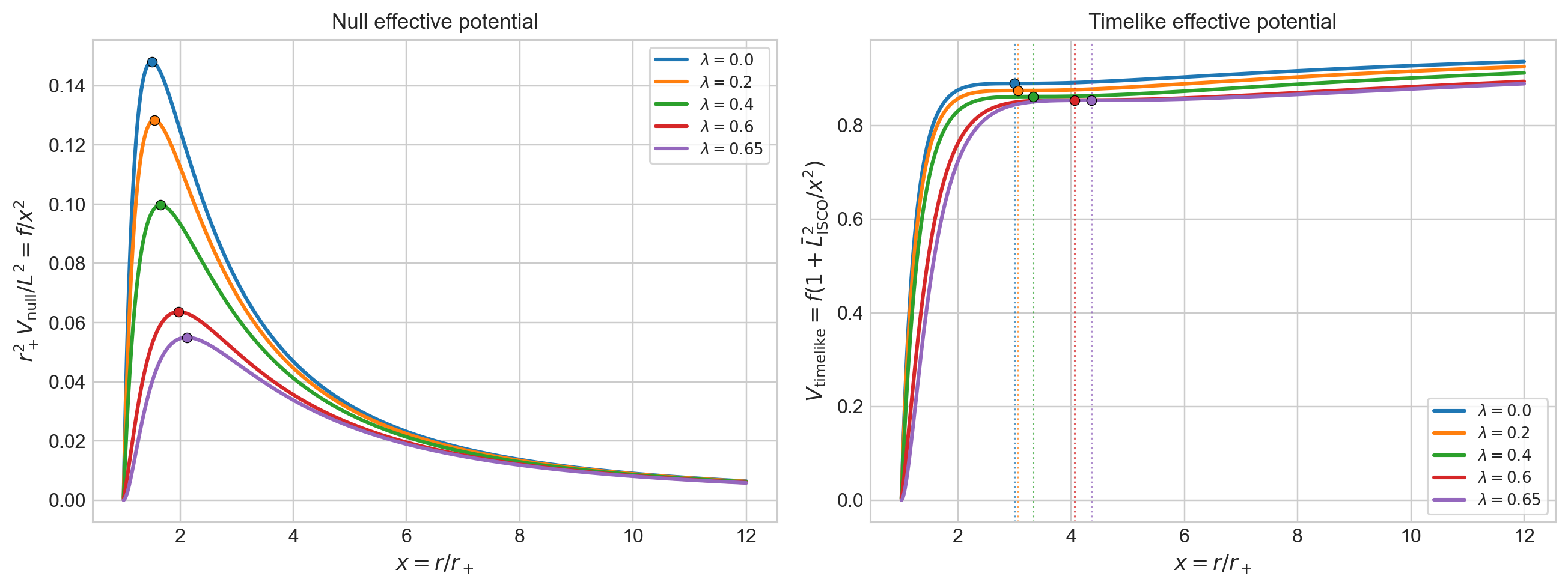}
\caption{Null and timelike effective potentials for the gravitational-self-energy regular black hole. \textbf{Left}: $f/x^2$, whose maximum gives the photon orbit. \textbf{Right}: $f(1+\bar L_{\rm ISCO}^2/x^2)$ for each benchmark value of $\lambda$, with $\bar L_{\rm ISCO}=L_{\rm ISCO}/r_+$. Larger $\lambda$ moves the relevant potential features outward.}
\label{fig:particle_effective_potentials}
\end{figure}

Figure~\ref{fig:particle_effective_potentials} shows the null effective potential $f/x^2$ and the timelike effective potential $f(1+\bar L_{\rm ISCO}^2/x^2)$, where $\bar L_{\rm ISCO}=L_{\rm ISCO}/r_+$ is chosen separately for each benchmark geometry. The null barrier maximum shifts outward and becomes broader as $\lambda$ increases, and the ISCO marker in the timelike potential also moves to larger radius. This behavior is quantified in Tables~\ref{tab:photon_particle} and~\ref{tab:isco_particle}. The photon-ring radius increases from the Schwarzschild value $x_{\rm ph}=1.5$ to $x_{\rm ph}=2.121$ at $\lambda=0.65$, while the critical impact parameter grows from $2.598r_+$ to $4.265r_+$.

The photon-ring angular frequency decreases monotonically, from $r_+\Omega_{\rm ph}=0.384900$ at $\lambda=0$ to $0.234469$ at $\lambda=0.65$. The Lyapunov exponent decreases even faster: the ratio $\Lambda_{\rm ph}/\Omega_{\rm ph}$ drops from unity in the Schwarzschild limit to $0.680$ at $\lambda=0.65$. Thus the gravitational self-energy correction makes the photon ring larger and less rapidly unstable in horizon units.

\begin{figure}[H]
\centering
\includegraphics[width=\textwidth]{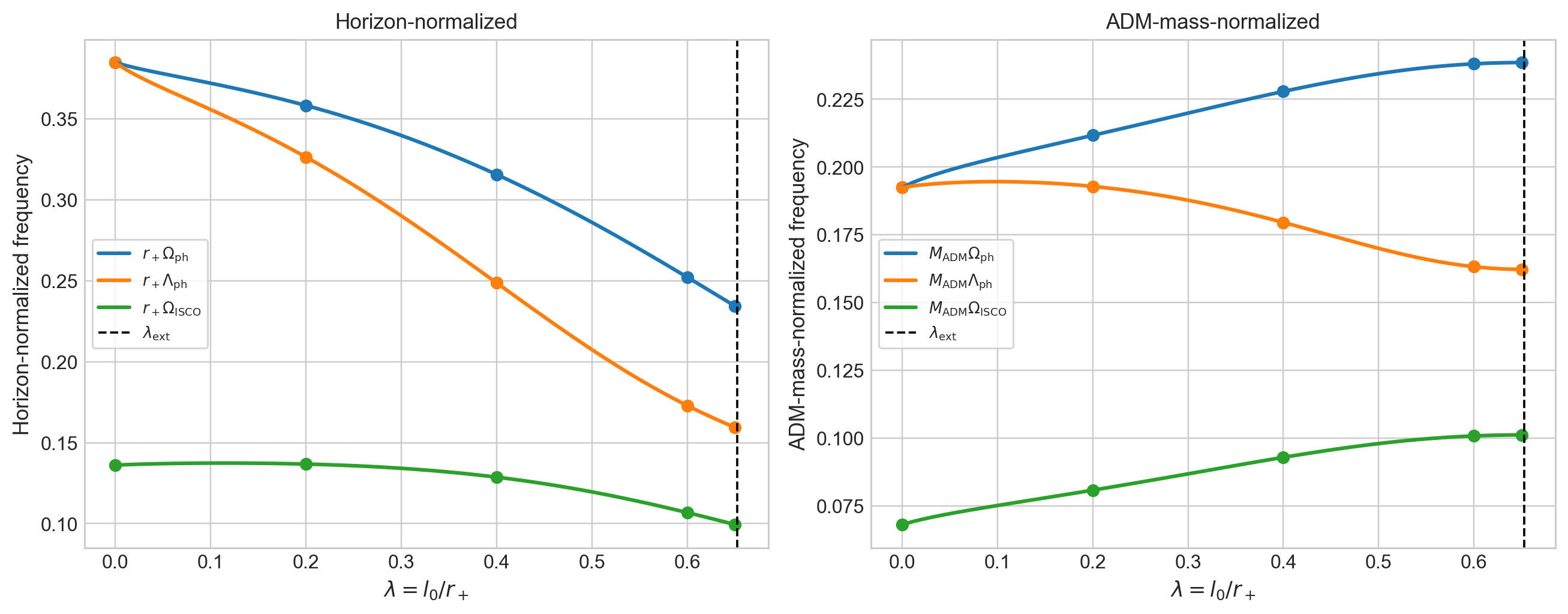}
\caption{Horizon-normalized (\textbf{left panel}) and mass-normalized (\textbf{right panel}) frequencies.}
\label{fig:particle_frequencies}
\end{figure}

It is useful to recall the standard eikonal connection between unstable null geodesics and quasinormal modes: in many single-barrier perturbation problems the real part of the large-multipole quasinormal frequency is proportional to the photon-ring orbital frequency, while the imaginary part is controlled by the null Lyapunov exponent, schematically $\mathrm{Re}\,\omega_{\ell n}\sim \ell\,\Omega_{\rm ph}$ and $\mathrm{Im}\,\omega_{\ell n}\sim-(n+1/2)\Lambda_{\rm ph}$ \cite{Cardoso2009Geodesic}. This correspondence gives a useful interpretation of the trends in Figure~\ref{fig:particle_frequencies}, but it is not a theorem for all backgrounds, spins or modified-gravity perturbation sectors; known exceptions and corrections should be kept in mind when using geodesic quantities as quasinormal-mode proxies \cite{Konoplya:2022gjp,Konoplya:2017wot,Bolokhov:2023dxq}.

\vspace{-36pt}

\begin{table}[H]
\centering
\caption{Photon-ring observables. The critical impact parameter controls the high-frequency shadow/capture scale, while $\Lambda_{\rm ph}$ is the null Lyapunov exponent.}
\label{tab:photon_particle}

\begin{tabular}{m{0.9in}<{\centering}m{0.9in}<{\centering}m{0.9in}<{\centering}m{0.9in}<{\centering}m{0.9in}<{\centering}m{0.8in}<{\centering}}
\toprule
$\boldsymbol{\lambda}$ & $\boldsymbol{x_{\rm ph}}$ & $\boldsymbol{b_{\rm ph}/r_+}$ & $\boldsymbol{r_+\Omega_{\rm ph}}$ & $\boldsymbol{r_+\Lambda_{\rm ph}}$ & $\boldsymbol{\Lambda_{\rm ph}/\Omega_{\rm ph}}$ \\
\midrule
        0.0 & 1.500000 & 2.598076 & 0.384900 & 0.384900 & 1.000 \\
        0.2 & 1.544470 & 2.790524 & 0.358356 & 0.326465 & 0.911 \\
        0.4 & 1.656820 & 3.167429 & 0.315713 & 0.248768 & 0.788 \\
        0.6 & 1.976102 & 3.963992 & 0.252271 & 0.172961 & 0.686 \\
        0.65 & 2.120580 & 4.264956 & 0.234469 & 0.159460 & 0.680 \\
        \bottomrule
\end{tabular}

\end{table}

\begin{table}[H]
\centering
\caption{Timelike ISCO observables. The efficiency is $100(1-E_{\rm ISCO})$.}
\label{tab:isco_particle}

\begin{tabular}{m{0.8in}<{\centering}m{0.9in}<{\centering}m{0.9in}<{\centering}m{0.9in}<{\centering}m{0.9in}<{\centering}m{0.9in}<{\centering}}
\toprule
$\boldsymbol{\lambda}$ & $\boldsymbol{x_{\rm ISCO}}$ & $\boldsymbol{E_{\rm ISCO}}$ & $\boldsymbol{L_{\rm ISCO}/r_+}$ & $\boldsymbol{r_+\Omega_{\rm ISCO}}$ & $\boldsymbol{\eta(\%)}$ \\
\midrule
        0.0 & 3.000000 & 0.942809 & 1.732051 & 0.136083 & 5.72 \\
        0.2 & 3.062056 & 0.934806 & 1.899853 & 0.136766 & 6.52 \\
        0.4 & 3.326025 & 0.928039 & 2.213704 & 0.128698 & 7.20 \\
        0.6 & 4.056169 & 0.923948 & 2.823786 & 0.106812 & 7.61 \\
        0.65 & 4.359178 & 0.923762 & 3.041064 & 0.099440 & 7.62 \\
        \bottomrule
\end{tabular}
\end{table}

\vspace{-12pt}

\vspace{-24pt}

\section{Shadow Radius}\label{sec5}

\begin{figure}[H]
\centering
\includegraphics[width=\textwidth]{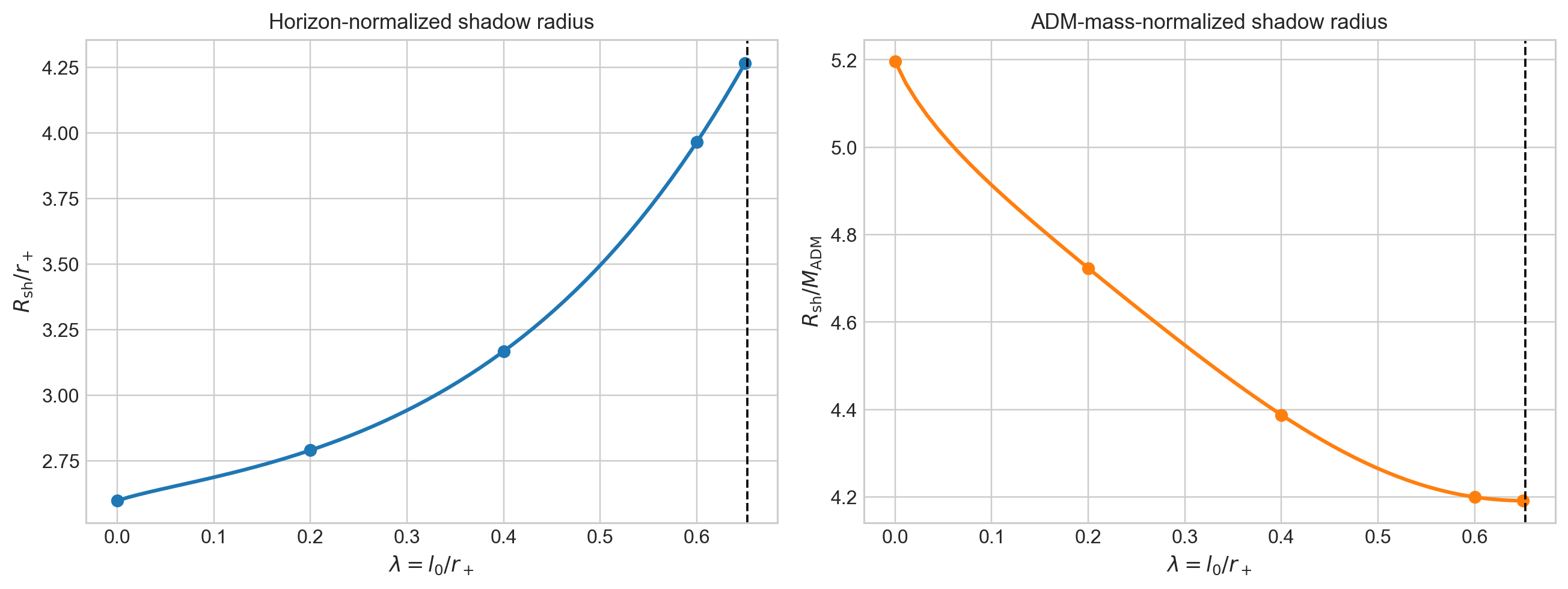}
\caption{Shadow radius of the gravitational-self-energy regular black hole. \textbf{Left}: $R_{\rm sh}/r_+$ grows monotonically with $\lambda$. \textbf{Right}: $R_{\rm sh}/M_{\rm ADM}$ decreases because the ADM mass increases relative to the fixed horizon radius. The dashed vertical line marks the extremal endpoint.}
\label{fig:particle_shadow_radius}
\end{figure}

The critical impact parameter in Table~\ref{tab:photon_particle} is also the shadow radius for a static observer at infinity. Figure~\ref{fig:particle_shadow_radius} and Table~\ref{tab:shadow_particle} show two useful normalizations. At fixed event-horizon radius, the shadow grows monotonically: $R_{\rm sh}/r_+$ rises from $2.598$ at $\lambda=0$ to $4.265$ at $\lambda=0.65$. The corresponding apparent area, proportional to $R_{\rm sh}^2$, is larger by a factor of $2.695$ at $\lambda=0.65$ than in the Schwarzschild limit with the same $r_+$.

The trend changes when the same radius is normalized by the ADM mass. Because the gravitational self-energy increases $M_{\rm ADM}/r_+$ along the branch, $R_{\rm sh}/M_{\rm ADM}$ decreases from $5.196$ in the Schwarzschild limit to $4.192$ at $\lambda=0.65$. Thus a horizon-scale comparison predicts a larger shadow, whereas an ADM-mass-normalized comparison predicts a smaller dimensionless shadow. This distinction is important for phenomenology: a model-independent angular shadow measurement constrains $R_{\rm sh}/D$, where $D$ is the distance from the observer to the black hole, so $R_{\rm sh}/D$ is the small-angle angular shadow radius, while converting it into a test of the metric also requires an independent mass and distance calibration.

The monotonic increase of $R_{\rm sh}/r_+$ follows directly from the outward displacement of the photon ring and the reduction of the photon-ring frequency, since $R_{\rm sh}/r_+=1/(r_+\Omega_{\rm ph})$. Near the remnant branch, therefore, the same null orbit that becomes less unstable also produces a larger horizon-normalized optical capture radius.

\vspace{-12pt}

\begin{table}[H]
\centering
\caption{Shadow-radius diagnostics. Here $R_{\rm sh}=b_{\rm ph}$ for an observer at infinity. The final two columns give the radius and area ratios relative to the Schwarzschild value at fixed $r_+$.}
\label{tab:shadow_particle}

\begin{tabular}{m{1.1in}<{\centering}m{1.1in}<{\centering}m{1.1in}<{\centering}m{1.1in}<{\centering}m{1.07in}<{\centering}}
\toprule
$\boldsymbol{\lambda}$ & $\boldsymbol{R_{\rm sh}/r_+}$ & $\boldsymbol{R_{\rm sh}/M_{\rm ADM}}$ & \textbf{Radius Ratio} & \textbf{Area Ratio} \\
\midrule
        0.0 & 2.598076 & 5.196152 & 1.000 & 1.000 \\
        0.2 & 2.790524 & 4.723427 & 1.074 & 1.154 \\
        0.4 & 3.167429 & 4.387891 & 1.219 & 1.486 \\
        0.6 & 3.963992 & 4.200047 & 1.526 & 2.328 \\
        0.65 & 4.264956 & 4.191604 & 1.642 & 2.695 \\
\bottomrule
\end{tabular}
\end{table}

\vspace{-24pt}

\section{Timelike Circular Orbits and ISCO}\label{sec6}

\begin{figure}[H]
\centering
\includegraphics[width=\textwidth]{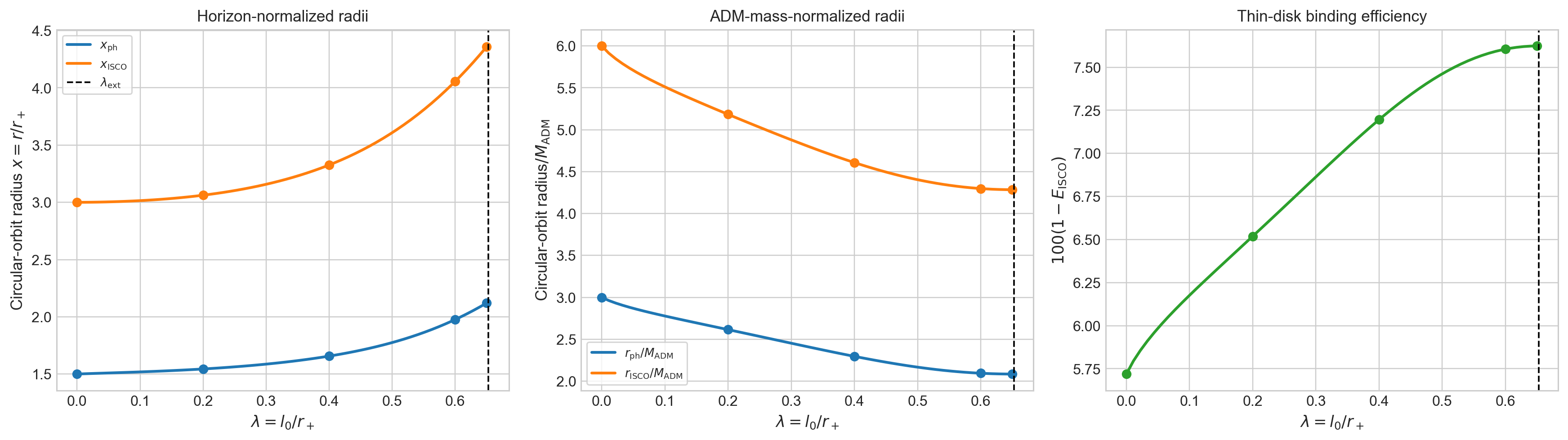}
\caption{Circular-orbit structure as a function of $\lambda=l_0/r_+$. \textbf{Left}: photon-ring and ISCO radii in the units of the event horizon radius. Central: photon-ring and ISCO radii in the units of the ADM mass.
\textbf{Right}: ISCO binding efficiency $100(1-E_{\rm ISCO})$. The dashed vertical line marks the extremal endpoint.}
\label{fig:particle_orbital_structure}
\end{figure}

The ISCO data are shown in Figure~\ref{fig:particle_orbital_structure} and Table~\ref{tab:isco_particle}. In unites of the event horizon the ISCO moves outward from $x_{\rm ISCO}=3$ in the Schwarzschild limit to $4.359$ at $\lambda=0.65$. The specific angular momentum increases over the same interval, reflecting the larger orbital radius, while the specific energy decreases from $E_{\rm ISCO}=0.942809$ to $0.923762$.

The binding efficiency $\eta=1-E_{\rm ISCO}$ therefore grows from $5.72\%$ to $7.62\%$. This increase should not be confused with a higher Hawking luminosity: it is a purely geodesic thin-disk efficiency, measuring the binding energy released by matter that reaches the ISCO. After a small rise at $\lambda=0.2$, the corresponding orbital frequency falls to $r_+\Omega_{\rm ISCO}=0.099440$ at $\lambda=0.65$, because the ISCO displacement outward dominates over the modest deepening of the binding energy.

\vspace{-24pt}

\section{Novikov-Thorne Flux Profile}\label{sec7}

As a simple accretion diagnostic, we compute the zero-torque Novikov-Thorne flux profile using the circular-orbit functions above. The assumed configuration is the standard idealized thin disk: gas is supplied at a constant mass accretion rate $\dot M$, lies on the equatorial plane, and moves through a sequence of nearly circular geodesics while viscosity transports angular momentum outward. The disk is taken to be geometrically thin, optically thick and radiatively efficient, so the dissipated binding energy is emitted locally from each annulus \cite{ShakuraSunyaev1973}. We neglect disk self-gravity, finite thickness, pressure corrections, magnetic stresses, returning radiation and spectral hardening, and we place the inner edge at the ISCO with a vanishing torque there. With $\bar L=L/r_+$ and $\bar\Omega=r_+\Omega$, the dimensionless flux per unit mass accretion rate is
\begin{equation}
\label{eq:nt_flux_particle}
 \frac{r_+^2\mathcal F(x)}{\dot M}
 =-\frac{\bar\Omega_{,x}}{4\pi x\left(E-\bar\Omega\bar L\right)^2}
 \int_{x_{\rm ISCO}}^x
 \left(E-\bar\Omega\bar L\right)\frac{d\bar L}{dx'}\,dx' .
\end{equation}

This is the standard Novikov-Thorne/Page-Thorne expression specialized to the present static, spherically symmetric metric \cite{NovikovThorne1973,PageThorne1974}. We use it only as a geometric comparison between backgrounds, not as a full radiative-transfer model.

\begin{figure}[H]
\centering
\includegraphics[width=\textwidth]{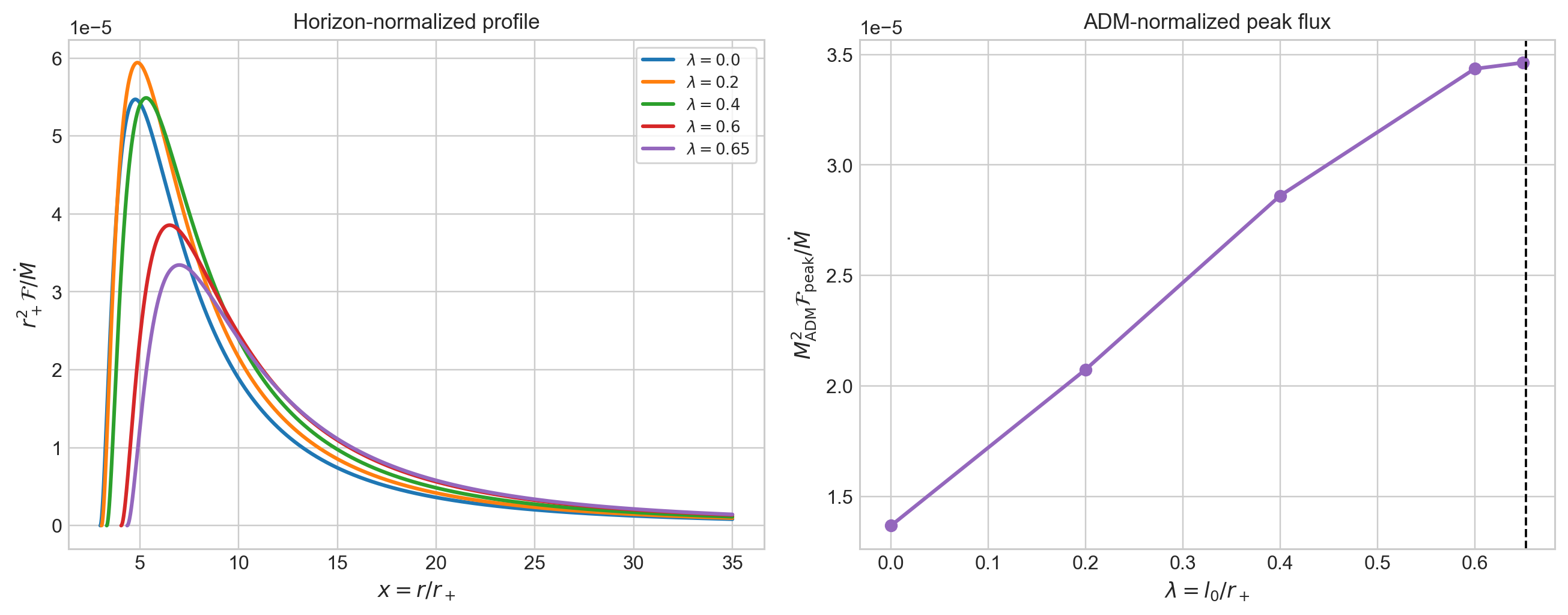}
\caption{Novikov-Thorne flux from one disk face profile for a zero-torque inner boundary at the ISCO. The peak shifts outward as $\lambda$ increases and is substantially reduced near $\lambda=0.65$.}
\label{fig:particle_nt_flux}
\end{figure}

The flux profiles are shown in Figure~\ref{fig:particle_nt_flux}, and the peak values are listed in Table~\ref{tab:nt_particle}. For $\lambda=0.2$ the peak is slightly larger than the Schwarzschild value, but the peak position has already moved outward. For $\lambda=0.4$ the peak height is nearly unchanged relative to Schwarzschild. By $\lambda=0.65$, very close to the remnant branch, the peak lies at $x_{\rm peak}=6.993$ and its height is reduced to $0.611$ of the Schwarzschild value. The dominant qualitative disk signature is therefore an outward displacement of the bright region, accompanied near extremality by a lower peak flux in horizon-normalized units.

\vspace{-36pt}

\begin{table}[H]
\centering
\caption{Peak of the dimensionless Novikov-Thorne flux profile. The final column is normalized to the Schwarzschild peak at $\lambda=0$.}
\label{tab:nt_particle}

\begin{tabular}{m{1.4in}<{\centering}m{1.4in}<{\centering}m{1.4in}<{\centering}m{1.4in}<{\centering}}
\toprule
$\boldsymbol{\lambda}$ & $\boldsymbol{x_{\rm peak}}$ & $\boldsymbol{r_+^2\mathcal F_{\rm peak}/\dot M}$ & \textbf{Peak Ratio} \\
\midrule
        0.0 & 4.775636 & 5.471 
 $\times$ 10\textsuperscript{$-$5} & 1.000 \\
        0.2 & 4.877826 & 5.942 $\times$ 10\textsuperscript{$-$5} & 1.086 \\
        0.4 & 5.328466 & 5.490 $\times$ 10\textsuperscript{$-$5} & 1.003 \\
        0.6 & 6.505031 & 3.856 $\times$ 10\textsuperscript{$-$5} & 0.705 \\
        0.65 & 6.993099 & 3.345 $\times$ 10\textsuperscript{$-$5} & 0.611 \\
        \bottomrule
\end{tabular}
\end{table}

\section{Conclusions}\label{sec8}

We have analyzed the optical, orbital and thin-disk diagnostics of a regular black hole sourced by gravitational self-energy. The calculation was performed on the black-hole branch parametrized by $\lambda=l_0/r_+$, with the outer horizon used as the primary length scale and the ADM mass used separately when discussing mass-normalized shadow observables.

The null-geodesic sector gives the clearest optical signature. As $\lambda$ increases, the photon ring moves outward, the critical impact parameter grows in horizon units, and both the photon-ring frequency and the null Lyapunov exponent decrease. Consequently the apparent shadow radius grows from $2.598r_+$ to $4.265r_+$ along the benchmark sequence, corresponding to a factor of $2.695$ increase in horizon-normalized shadow area. The same data look different when normalized by $M_{\rm ADM}$: because the self-energy contribution increases the ADM mass relative to the fixed horizon radius, $R_{\rm sh}/M_{\rm ADM}$ decreases from $5.196$ to $4.192$. This scale dependence is important for any phenomenological comparison with angular shadow measurements.

The timelike circular-orbit sector shows a parallel outward displacement. The ISCO moves from $3r_+$ in the Schwarzschild limit to $4.359r_+$ at the near-extremal benchmark, while the binding efficiency increases from $5.72\%$ to $7.62\%$. The ISCO frequency is then lower than in the Schwarzschild limit, because the increase in the orbital radius dominates. In the eikonal interpretation, the decrease of $\Omega_{\rm ph}$ and $\Lambda_{\rm ph}$ would suggest lower real frequencies and weaker damping for the corresponding large-multipole quasinormal modes, but the caveat above stresses that this geodesic correspondence has known limitations and must not replace a direct perturbation calculation in exceptional cases.

The Novikov-Thorne estimate translates these geodesic trends into an accretion diagnostic under a number of assumptions: zero-torque ISCO boundary condition, absence of magnetic stresses, no returning radiation, no spectral hardening, and no radiative-transfer effects. The flux maximum shifts outward for every nonzero deformation, and near the remnant branch its height is reduced to about $61\%$ of the Schwarzschild peak in horizon-normalized units. Taken together, the outward-moving photon ring, the scale-dependent shadow radius, the displaced ISCO and the modified disk-flux profile provide a consistent set of signatures of the gravitational-self-energy regularization. Future work should test these geodesic indicators against full perturbation spectra, radiative-transfer images and dynamical accretion models.

		\section*{Funding}
This research received no external funding.
 
		\section*{Institutional Review Board Statement}
Not applicable.

		\section*{Informed Consent Statement}
Not applicable.

		\section*{Data Availability Statement}
Not applicable.

		\section*{Conflicts of Interest}
		
The authors declare no conflict of interest. 
	
	\section*{Use of AI and AI-Assisted Technologies}
	
During the preparation of this manuscript, the author employed ChatGPT (GPT-5, OpenAI) to assist in the refinement of language and improvement of textual clarity and style. After using this tool/service, the author reviewed and edited the content as needed and takes full responsibility for the content of the published article.

\end{document}